\documentclass{aa} 

\usepackage{graphicx,natbib,url,amssymb}
\bibpunct{(}{)}{;}{a}{}{,} 
\usepackage{txfonts}
\newcommand{\msun}{M\ensuremath{_\odot}}	
\newcommand{\mdot}{\ensuremath{\dot M}}		
\newcommand{\sgra}{Sgr A*}

\newcommand{\bh}{black hole}

\newcommand{\urlfn}[1]{\footnote{\url{#1}}}

\begin{document}
\title{A time-dependent jet model for the emission from Sagittarius A*}
\author{
	Dipankar Maitra\inst{1} \and 
	Sera Markoff\inst{1} \and 
	Heino Falcke\inst{2,3} 
}
\offprints{D.\ Maitra, \email{D.Maitra@uva.nl}} 
\institute{
Astronomical Institute ``Anton Pannekoek'', University of Amsterdam, 
Science Park 904, 1098 XH Amsterdam, The Netherlands 
\and 
Department of Astrophysics, Institute for Mathematics, Astrophysics and 
	Particle Physics, Radboud University, P.O. Box 9010, 6500 GL Nijmegen, 
	The Netherlands 
\and 
ASTRON, Oude Hoogeveensedijk 4, 7991 PD Dwingeloo, The Netherlands 
}

\date{} 

\abstract
{The source of emission from \sgra, the supermassive \bh\ at the Galactic 
  Center, is still unknown.  
  Flares and data from multiwavelength campaigns 
  provide important clues about the nature of \sgra\ itself.
}
{Here we attempt to constrain the physical origin of the broadband 
 emission and the radio flares from \sgra.}
{We developed a time-dependent jet model, which for the first time 
  allows one to compare the model predictions with flare data from \sgra.
  Taking into account relevant cooling mechanisms, we calculate the 
  frequency-dependent time lags and photosphere size expected in the 
  jet model. The predicted lags and sizes are then compared with 
  recent observations.
}
{Both the observed time lags and size-frequency relationships are 
  reproduced well by the model.  The combined
  timing and structural information strongly constrain the speed of the 
  outflow to be mildly relativistic, and the radio flares are likely to be
  caused by a transient increase in the matter channelled into the jets.  The 
  model also predicts light curves and structural information at other 
  wavelengths which could be tested by observations in the near future.
}
{We show that a time-dependent relativistic jet 
  model can successfully reproduce: (1) the quiescent broadband spectral energy 
  distribution of \sgra, (2) the observed 22 and 43 GHz light curve 
  morphologies and time lags, and (3) the frequency-size relationship. 
  The results suggest that the observed emission at radio frequencies from 
  \sgra\ is most easily explained by a stratified, optically thick, mildly 
  relativistic jet outflow. 
  Frequency-dependent measurements of time-lags and intrinsic source size 
  provide strong constraints on the bulk motion of the jet plasma.
}

\authorrunning{Maitra, Markoff, Falcke}
\titlerunning{A Time-Dependent Jet Model for Sgr A*}

\maketitle

\section{Introduction}
\label{s:intro}
The compact radio source Sagittarius A* (hereafter \sgra), at an
estimated distance of $8.4\pm0.6$ kpc \citep{ReidMentenZheng2009}
and mass of $\sim 4\times10^6\;\msun$ \citep{GhezSalimWeinberg2008,
GillessenEisenhauerTrippe2009}, is our closest supermassive \bh.
When compared to other active galactic nuclei (AGN), \sgra\
is extremely under-luminous at all wavelengths, suggesting either
that the mass accretion rate (\mdot) onto the \bh, or the radiative
efficiency, is very low.  Polarization measurements suggest that
$\mdot\leq 10^{-7}M_\odot/{\rm yr}$ \citep{MarroneMoranZhao2007}.
Recent multiwavelength campaigns that have monitored \sgra\ 
simultaneously
from radio to X-rays have found that even though the luminosity
of \sgra\ is low relative to its AGN counterparts, its emission
is quite variable at all wavelengths.  Flaring activity at both 
sub-mm
and higher frequencies has been observed on timescales ranging from
minutes to hours, suggesting that the emission at these frequencies is
produced very close to the supermassive \bh, probably within a few tens of
gravitational radii ($R_g$=$GM/c^2$). Although flaring appears to be nearly
simultaneous at these frequencies \citep{MarroneBaganoffMorris2008,
EckartSchoedelGarcia2008,Dodds-EdenPorquetTrap2009},
the flares seem to have an increasingly long time-delay at longer wavelengths.
For example, \citet{Yusef-ZadehWardleHeinke2008} found that the 43 GHz
flares precede the 22 GHz flares by $20\pm10$ minutes, and that 
350 GHz flares precede those of 43 GHz by $30-60$ minutes.

The physical origin of the observed electromagnetic radiation
from \sgra\ has been under active debate for more than a decade.
Attempts to model the emission from \sgra\ can be broadly
classified into two categories: (1) accretion inflow models
\citep{Melia1992, NarayanMahadevanGrindlay1998,LiuMelia2002,
YuanShenHuang2006}, and (2) relativistic outflow/jet models
\citep{FalckeBiermann1999,FalckeMarkoff2000,YuanMarkoffFalcke2002}.
The radio, NIR, and X-ray flare light curves were modeled
assuming expansion of a spherical plasma blob \citep[see
e.g.,][]{Yusef-ZadehWardleHeinke2008, EckartBaganoffMorris2009}.
Recently \citet{FalckeMarkoffBower2009} suggested that the
radio time-lag data and measurements of intrinsic size at various
frequencies \citep[e.g.,][]{BowerFalckeHerrnstein2004,ShenLoLiang2005,
DoelemanWeintroubRogers2008} can be used to resolve the degeneracy between
inflow and outflow models.  It has long been suspected that radio flares
such as those reported by \citet{Yusef-ZadehWardleHeinke2008} could be
``smoking gun evidence'' of jets in sources such as
\sgra\ \citep{Falcke1999b}.

While jets have not been directly imaged in \sgra, several
observations strongly suggest the presence of jets in this source.
The flat/slightly inverted radio spectral energy distribution (SED)
\citep[signature of a compact self-absorbed jet in the radio; see
e.g.,][] {BlandfordKonigl1979} observed from \sgra\ are similar to 
those of other
low-luminosity AGNs.  M81 and \sgra\ have very similar spectra and
polarization properties in the cm-radio band, and M81 has been found to have
weak
jets \citep{NagarFalckeWilson2005}.  Radio observations of the stellar
X-ray binary system A0620$-$00 \citep{GalloFenderMiller-Jones2006} at
almost near quiescent luminosity suggest that compact jets are present
at bolometric luminosities as low as $\sim$10$^{-7}$ L$_{\rm Edd}$.
If jet physics and accretion flow scale with the mass of the \bh, then \sgra\
(L$_{\rm bol}$$\sim$10$^{-9}$ L$_{\rm Edd}$) is also expected to harbor
a faint, compact jet.  Steady-state jet models have previously been
used to model the broadband quiescent spectra of \sgra\ successfully
\citep{FalckeMarkoff2000, YuanMarkoffFalcke2002}; steady-state models,
however, cannot predict the flaring activity.

In this paper we develop a proper treatment of the time-dependent
cooling processes of the particle distribution, while keeping the
basic hydrodynamic outflow model the same as in \citet{Falcke1996},
allowing a semi-analytical treatment of the problem.  Once dominant
cooling mechanisms are taken into account, the time-dependent jet model
naturally accounts for all the observational constraints, namely, (a) the
broadband SED, (b) the observed light-curve morphologies and time-lags,
and (c) the observed frequency--size relationship.

\section{Model}
\label{s:jetmodel}
We assume that a fraction of the accretion inflow is chanelled into
two symmetric, collimated supersonic outflows.  The energy distribution
of the radiating leptons entering the jet is assumed to be a thermal,
relativistic Maxwell-Juttner distribution.  Most of the kinetic energy of
the jet is assumed to be carried by cold, non-radiating baryons.  The bulk
speed at the base of the jet is assumed to be the sound speed, and beyond
the base (or the ``nozzle'') the jet plasma accelerates longitudinally
via pressure gradient and expands laterally with the sound speed
$\beta_s \sim 0.4$.  We adopt the standard one-fluid approximation used
in magnetohydrodynamics \citep[MHD; see e.g.][]{Leismannetal2005}, and
treat the magnetized jet plasma as a single-component fluid of adiabatic
index $4/3$.  The longitudinal velocity profile is obtained by solving the
relativistic Euler equation, and the magnetic field strength estimated
via flux conservation \citep[see][for details of the hydrodynamics of
a pressure driven supersonic jet]{Falcke1996,FalckeMarkoff2000}.

The observed spectrum from the compact jet is modeled as the sum of emission 
from cylindrical segments along the jet axis. 
Within each segment we consider the following processes influencing the 
local evolution of the particle distribution:
\begin{itemize}
 \item {\em Synchrotron cooling:}
   Losses due to synchrotron emission are given by
    $\dot\gamma_{_{Syn}}$=$4c\sigma_{_T} U_B \gamma^2/(3m_e c^2)$,
   where $\sigma_{_T}$ is the Thomson cross-section and 
   $U_B$=$B^2/8\pi$ is the magnetic energy density.

 \item {\em Inverse Compton (IC) cooling:}
   Losses due to IC scatterings are given by
    $\dot\gamma_{_{Com}}$=$4c\big<\sigma_{_{KN}}U_{rad}(t)\big>_{\gamma}\gamma^2/
   (3m_e c^2)$,
   where, following \citet{dekool1989} we define 
    $\big<\sigma_{_{KN}}U_{rad}(t)\big>_{\gamma}=4\pi/c
    \int \sigma_{_{KN}}(\gamma,\nu) \; J_{\nu} \; d\nu$.
   Here $\sigma_{_{KN}}(\nu,\gamma)$ is the Klein-Nishina correction to 
   the scattering cross-section and $J_{\nu}$ the mean intensity. 

 \item {\em Adiabatic expansion:} 
   Cooling due to adiabatic expansion is given by
   $\dot\gamma_{_{ad}}$=$(1/3) \gamma \; \nabla\cdot{\vec{v}}$,
   where $\nabla\cdot{\vec{v}}$ denotes the divergence of the 
   bulk velocity field \citep{Longair1992}.
\end{itemize}

For numerical discretization we assume that the plasma in each segment 
cools for the time it requires to cross the segment.
We follow the evolution of such a ``parcel'' of jet plasma as it moves from one
segment to another, i.e. in a comoving frame, without any particle loss/escape.
For simplicity we assume that this comoving ``parcel'' of particles cools and 
radiates independently, i.e. is not influenced by its neighbors.
While this is obviously a simplification, given the overall simplicity of the 
model we feel this is justified.
The continuity equation for the time evolution of leptons in this case is 
given by
\begin{equation}
\frac{\partial N(\gamma,t)}{\partial t} = \frac{\partial}{\partial\gamma} 
\left[ \dot\gamma(\gamma,t) N(\gamma,t)\right]
\;\; ; \; \dot\gamma(\gamma,t) = \dot\gamma_{_{Syn}}+\dot\gamma_{_{Com}}+
\dot\gamma_{_{ad}}
\label{eq:continuity}
\end{equation}

Given the low luminosity of \sgra\ ($\sim$10$^{-9}$ L$_{\rm Edd}$), the
compactness parameter \citep[$\ell$$\equiv$L$\sigma_{T}$/(Rm$_e$c$^3$);
see][]{GuilbertFabianRees1983} is very small ($\ell$$\sim$10$^{-6}$
for R=10 R$_g$).  Therefore pair processes are not important
here.  Also, there is no injection term in (\ref{eq:continuity})
since particle acceleration seems to be very weak or lacking in
\sgra\ \citep{Markoff2005}.  Recasting (\ref{eq:continuity}) as a
Fokker-Planck equation, we used the fully implicit Chang-Cooper algorithm
\citep{cc1970,cg1999} to solve it.

Once the time-evolved particle distribution solution is obtained, we
then compute the emitted SED due to angle averaged synchrotron emission
using eq.(10) of \citet{cs1986} for relativistic electrons with Lorentz
factor $\geq$2, and using eq. (8) of \citet{kat2006} for electrons in
the cyclo-synchrotron regime with Lorentz factor $<$2.
IC is computed incorporating the Klein-Nishina correction for scattering
beyond Thomson limit following the prescription of \citet{bg1970}. The
seed photons for IC are the photons produced locally by synchrotron
emission (synchrotron self-Compton; SSC).  In the absence of detailed
spatial information of the hot stars that constitue the central star
cluster close to \sgra\ \citep[see e.g.][] {Genzeletal2000}, we estimated
emission from these stars by assuming 100 stars of temperature 10,000 K
each, situated 0.5 parsecs from \sgra.  The photon density of locally
produced synchrotron+SSC photons in the jet is at least 3 orders of
magnitude higher than that due to total photon density from these stars,
suggesting that external inverse Compton losses due to photons from the
nearby cluster is not important.

Since the bulk motion of the emitting jet plasma is relativistic
we first compute the Doppler factor $\mathcal{D}=[\Gamma(1-\beta
cos\theta)]^{-1}$, where $\theta$ is the angle between velocity and line
of sight, ${c\bf\beta}$ is the bulk speed, and $\Gamma=1/\sqrt{1-\beta^2}$
is the bulk Lorentz factor.  Then the observed SED is calculated by
applying the appropriate special relativistic transformations of the
emitted frequency and flux \citep[see e.g.][]{LindBlandford1985}, and
taking into account time-lags due to the spatial extent of the flow.

For any given frequency, when the flux from each jet segment is plotted as
a function of distance from the \bh, it roughly resembles a bell-shaped
curve (in the self-absorbed part of the jet). The full width at half
maximum (FWHM) is an indicator of the size of the photosphere of the jet
at this frequency. We used this procedure to calculate the frequency--size
relationship.

In our model the radio flares are caused by a perturbation in \mdot\ at 
the jet-base, leading to a transient density enhancement.
This density enhancement during a flare of duration $t_d$ is 
quantified by the parameter $f_n$ so that the number density at 
the base ($n_0$) as a function of time is given by
\begin{equation}
n_0(t) = \left\{ \begin{array}{ll}
             (1+f_n)\; n_{0} & \mbox{$(0 \leq t \leq t_d)$} \\
             n_{0}         & \mbox{otherwise.} \\
	     \end{array}
	 \right.
\label{eq:numden}
\end{equation}

As the overdensity propagates outward along the jet, the additional
pressure in the overdense region would cool and expand into the
neighboring segments with lower pressure and density.  While a full
computation of this diffusion (and resultant cooling) would require
MHD simulations beyond the scope of the present paper, the effect
can still be captured by assuming an additional velocity dependent
cooling term in the continuity equation for the overdense region.
Assuming that the overdense region expands longitudinally into the
neighboring segments with some fraction of the sound speed, we write
this additional cooling term as a function of distance from the 
\bh\ ($z$) as $\dot \gamma / \gamma = f_o \beta_s c/ z$.  Thus $f_o
\beta_s c$ can be interpreted as the longitudinal speed with which
the overdense region expands.

\section{Modeling the quiescent broadband emission and the radio flares 
from \sgra on July 17, 2006}
\label{s:initialpars}
We test the model against the simultaneous observations of \sgra\
in 22 and 43 GHz on 2006 July 17 \citep{Yusef-ZadehWardleHeinke2008},
which has the best simultaneous coverage to date of a flare in both
frequencies.  Indirect evidence suggests that \sgra\ is seen under large
inclination angle \citep{MarkoffBowerFalcke2007,FalckeMarkoffBower2009}.
For simplicity we therefore assume that the jets are perpendicular to our
line-of-sight, even though the model can handle any inclination angle.
The free parameters of the model are: temperature ($T_e$) and density
($n_0$) of the thermal leptons at the base, location of the sonic point or
the nozzle ($h_0$), radius of the nozzle ($r_0$) and its magnetic field
($B_0$).  The flare is parametrized by its start time ($t_0$), duration
($t_d$), density enhancement fraction ($f_n$) and expansion speed of
the overdense region ($f_{o}$, as a fraction of the sound speed).

\begin{figure}
\resizebox{\hsize}{!}
{\includegraphics[angle=-90]{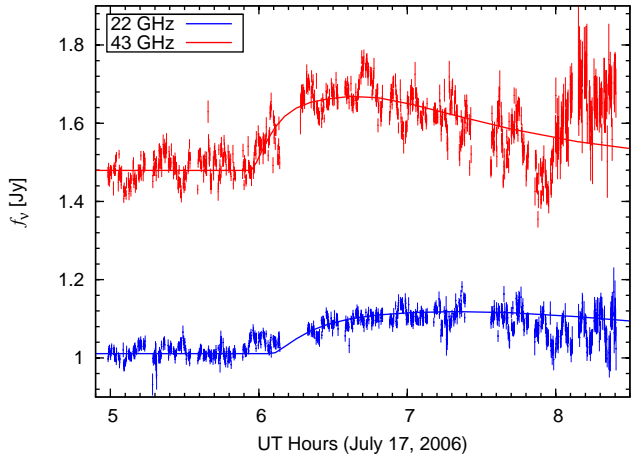}}
\caption{\label{f:lc_43_22} Comparison of model with data for the flare on
2006 July 17 at 43 and 22 GHz. The 43 GHz data with errorbars \citep[from][]
{Yusef-ZadehWardleHeinke2008} are shown in red, and 22 GHz data+model in 
blue. As noted in \S\ref{s:initialpars}, we model the flare which created 
a peak in the 43 GHz light curve near 6.5 hours UT.
}
\end{figure}

\begin{figure}
\resizebox{\hsize}{!}
{\includegraphics[angle=-90]{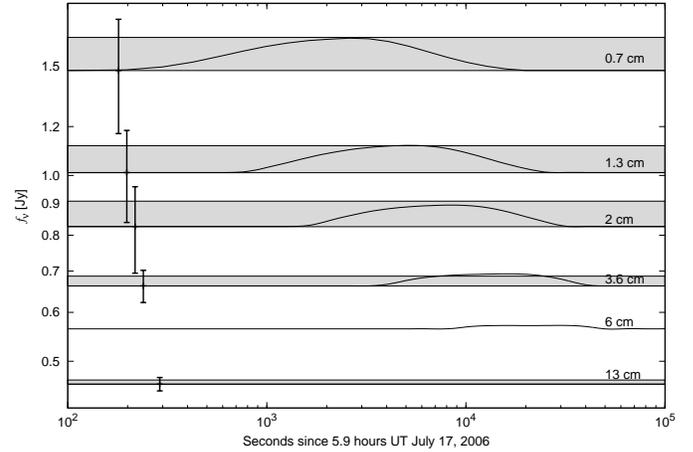}}
\caption{\label{f:lc} Model predicted light curves of \sgra\ at key
  radio frequencies using model parameters shown in
  Table~\ref{t:fitpars}.  Also shown as error bars between t=200--300s 
  are the rms flux variability of \sgra\ averaged over many flares as
  compiled by \citet{FalckeMarkoffBower2009}. 
  The gray bands have a width of 0.6 times the rms variability of \sgra, 
  which is roughly matched to the specific flare under consideration.
}
\end{figure}

The good agreement of the data with model light curves is shown
in Fig.~\ref{f:lc_43_22}.  The model parameters are given in
Table~\ref{t:fitpars}.  Note that we only model the flare which peaks
around 6.5 hours UT in 43 GHz.  The increase in flux in both bands
during the end of the observations may be the beginning of another
flare as suggested by \citet{Yusef-ZadehWardleHeinke2008}, but we do
not model this due to lack of full coverage.  In Fig.~\ref{f:lc} we
show model predicted light curves at 0.7, 1.3, 2, 3.6, 6 and 13 cm.
From Fig.~\ref{f:lc_43_22} it is clear that the light curves are
asymmetric and even a phenomenological model would require (a) rise
time-scale, (b) decay time-scale, and (c) amplitude. In our model also
three parmeters describe the flare completely, viz. $t_d$ (which can be
roughly associated with the rise rime), $f_o$ (associated with decay
time), and $f_n$ (associated with amplitude).  It is thus important
that a single set of flare parameters fits two rather disparate light
curves at two frequencies.  Moreover, the decrease of amplitude is
consistent with the decrease in average variability with frequency.
A compilation of the rms flux variability of \sgra\ averaged over
many flares by \citet{FalckeMarkoffBower2009} from VLA and GBI data
\citep{Falcke1999a,HerrnsteinZhaoBower2004} shows that the rms variability
decreases with increasing wavelength. The relative amplitude of this flare
at both 43 and 22 GHz is about 0.6 times the long-term rms variability,
therefore this flare was not extraordinarily bright or faint, and may
be considered as a fairly typical flare from \sgra.

\begin{figure}
\resizebox{\hsize}{!}
{\includegraphics[angle=-90]{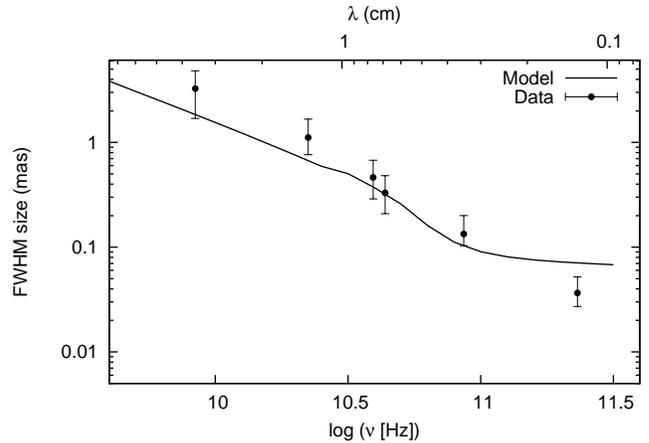}}
\caption{\label{f:fs} Comparison of model predicted frequency-size 
relationship (solid line) with observations. The data are from 
\citet{BowerFalckeHerrnstein2004,ShenLoLiang2005,
DoelemanWeintroubRogers2008}, which were compiled and corrected for 
scattering by \citet{FalckeMarkoffBower2009}.}
\end{figure}

The frequency--size relationship predicted by the model is displayed
in Fig.~\ref{f:fs}, showing reasonable agreement with the data at
low frequencies where the jet becomes optically thick. The discrepancy
at shorter wavelengths where the spectrum is optically thin is
expected given that our model ignores general relativistic effects,
and simplifies the physics of jet formation and launching, both of
which must play an important role very close to the \bh.  For this
model, the rate at which matter is fed to both jets is $\dot M = 2
n_0 m_p c \beta_s \gamma_s \pi r_0^2 = 2.4 \times 10^{-9}$ $\msun$
yr$^{-1}$, which is about 2 orders of magnitude lower than the upper
limit of $10^{-7}$ $\msun$ yr$^{-1}$ suggested by linear polarization
observations.

\section{Discussion}
\label{s:summary}
In this paper we have shown that a time-dependent relativistic jet model 
can explain most of the observational features seen in \sgra.  
The main conclusions of this work are:
\begin{itemize}
\item The model presented here can describe the quiescent
      broad band SED of \sgra, as well as its long-term radio variability.
\item Assuming that the radio flares are caused by a transient density 
      enhancement at the base of the jet, which then propagates outward, 
      the model can also fit the 22 and 43 GHz light curve
      morphology of the flares seen on 2006 July 17 by 
      \citet{Yusef-ZadehWardleHeinke2008}.  We predict light curve morphology
      and delays at longer wavelengths, which can be used to test the model
      with future observations.
\item The model predicted frequency--size relationship also matches quite 
      well with the observed radio data.
\end{itemize}

The radio emission in our model originates in the self-absorbed,
optically thick part of the outflow, where adiabatic losses dominate
over radiative cooling. Therefore variability in the radio light curves
largely traces the expansion of the jet plasma. Assuming that the flares
have the same frequency--size relationship as the quiescent emission,
the time delay measurements combined with frequency--size measurements
strongly constrain the bulk speed of the jet plasma in this model and
rule out subrelativistic expansion speeds.  An alternate model used by
\citet{Yusef-ZadehWardleHeinke2008} to fit the radio light curves assumes
adiabatic, subrelativistic expansion of a spherical blob of plasma,
i.e. the classic \citet{vanderLaan1966} model.  Our modeling, at the very
least, shows that the van der Laan model is not unique and in fact the jet
model, in contrast to the van der Laan model, not only correctly predicts
the light curves and the overall variability amplitudes, but also fits
the frequency--size relation without violating any other observational
constraints.  It must be kept in mind that the jet model does not attempt
to explain the NIR or X-ray flares.  The NIR/X-ray flares trace particle
(re)energization and cooling very close to the \bh, while only marginally
affecting the optically thick radio flux \citep{MarkoffFalckeYuan2001},
and no tight correlation between radio and X-ray flares have been found
so far.  A full general relativistic MHD model including radiative cooling
\citep[see e.g.][]{FragileMeier2009,MoscibrodzkaGammieDolence2009} would
be required to model the jet launching region.  However the simple model
presented here captures the important physics and shows that the jet
model explains the radio flaring properties naturally, due to simple
adiabatic expansion of overdensities in the outflow (likely linked to
variations in the accretion rate).  Future coordinated multiwavelength
campaigns, especially measurements of time-lags, sizes and positional
offset at other wavelengths, will enable a better understanding of the
velocity profile of the jet from \sgra.

\begin{acknowledgements}
We would like to thank Farhad Yusef-Zadeh for providing us the data of
the radio light curves, and an anonymous referee for valuable comments.
\end{acknowledgements}

\begin{table}
\caption{
Model parameters for the flare of 2006 July 17.  The mass ($M_{\rm
BH}$) and distance were taken from their latest estimates \citep[see
e.g.][]{GillessenEisenhauerTrippe2009,ReidMentenZheng2009}.  Inclination
of the jet axis to the line of sight was fixed to 90 degrees.  The base
of the jet is parametrized by radius of the nozzle ($r_0$), location of
the sonic point ($h_0$), magnetic field strength ($B_0$), number density
($n_0$) and temperature ($T_e$) of the thermal leptons.  The flare
is characterized by its start time ($t_0$), duration ($t_d$), density
enhancement fraction ($f_n$), expansion speed of the overdense region
($f_{o}$, as a fraction of the sound speed). Parameters marked with an
asterisk ($^*$) were not varied during fitting.
}
\label{t:fitpars}
\centering
\begin{tabular}{l l l l}
\hline\hline
Srl. No. & Parameter	& Value & Unit                  \\
\hline
1  & $M_{\rm BH}^*$  & 4	& $10^6$ M$_{\odot}$	\\
2  & Distance$^*$    & 8.4	& kpc			\\
3  & Inclination$^*$ & 90	& degrees		\\
4  & $r_0$           & 7	& R$_{\rm g}$		\\
5  & $h_0$           & 9	& R$_{\rm g}$		\\
6  & $B_0$           & 147	& Gauss			\\
7  & $n_0$           & 7	& $10^4$ cm$^{\rm -3}$	\\
8  & $T_e$           & 7	& $10^{10}$ K		\\
9  & $t_0$           & 5.90	& hours UT              \\
10 & $t_d$	     & 1800	& seconds               \\
11 & $f_{n}$         & 0.60	&			\\
12 & $f_{o}$         & 0.12	&			\\
\hline
\end{tabular}
\end{table}

\bibliographystyle{aa} \bibliography{dmrefs}
\end{document}